\documentclass[pdflatex,sn-nature]{sn-jnl}

\usepackage{graphicx}%
\usepackage{multirow}%
\usepackage{amsmath,amssymb,amsfonts}%
\usepackage{amsthm}%
\usepackage{mathrsfs}%
\usepackage[title]{appendix}%
\usepackage{xcolor}%
\usepackage{textcomp}%
\usepackage{manyfoot}%
\usepackage{booktabs}%
\usepackage{algorithm}%
\usepackage{algorithmicx}%
\usepackage{algpseudocode}%
\usepackage{listings}%
\usepackage{siunitx}
\theoremstyle{thmstyleone}%
\theoremstyle{thmstyletwo}%

\theoremstyle{thmstylethree}%

\begin{document}
\title{Active spinners drive passive matter into chiral rotors}









\setlength {\marginparwidth }{2cm}
\newcommand{\ans}[1]{\textcolor{teal}{#1}}
\newcommand{\ds}[2]{\textcolor{red}{#1}}
\newcommand{\Rao}[2]{\textcolor{magenta}{#1}}

\def\ds#1{\textcolor{blue}{(Dennis: #1)}}


\author[1]{\fnm{Dennis} \sur{Schorn}}
\equalcont{These authors contributed equally to this work.}

\author[2]{Anpuj Nair S}
\equalcont{These authors contributed equally to this work.}

\author[2]{Stijn van der Ham}

\author*[1]{Benno Liebchen}\email{benno.liebchen@pkm.tu-darmstadt.de}

\author*[2]{Hanumantha Rao Vutukuri}\email{h.r.vutukuri@utwente.nl}

\affil[1]{Institut für Physik der Kondensierten Materie, Technische Universität Darmstadt, Hochschulstraße 8, D-64289 Darmstadt, Germany}
\affil[2]{Active Soft Matter and Bio-inspired Materials Lab, Faculty of Science and Technology, MESA+ Institute for Nanotechnology and Center for Brain-Inspired Nano Systems, University of Twente, Enschede, Netherlands}

\date{July 29. 2026}


\abstract{Active spinners inject angular momentum into their surroundings without persistent translation, providing a route to chiral active matter distinct from self-propelled particles moving along curved trajectories. Recent studies have shown that spinner fluids can self-organize into rotating clusters with circulating edge currents characteristic of odd-viscous fluids. Whether angular-momentum injection alone can transmit chirality and edge-current dynamics to ordinary passive matter—and thus serve as a generic mechanism for imparting chiral functionality to undriven material components—remains an open question.
Here, we show experimentally and numerically that purely rotational magnetic spinners can transfer chirality and edge-current dynamics to passive colloidal matter. At high passive-particle fraction, active spinners assemble into rotating clusters surrounded by passive matter. Remarkably, upon changing composition and spinner attraction, this organization turns inside out: passive colloids form rotating clusters surrounded by a chiral spinner fluid. These passive rotors exhibit edge currents and the same angular-velocity scaling, {\unboldmath $|\Omega|\sim R^{-2}$}, as their active counterparts. A minimal model shows that spinner-mediated transverse and non-reciprocal interactions are sufficient to reproduce these complementary chiral states and their boundary-driven dynamics.
Our results establish angular-momentum injection by active spinners as a route to endow otherwise passive matter with chirality, collective rotation and boundary transport.}


\maketitle

\section*{Introduction}

Chiral active matter comprises systems whose microscopic constituents continuously consume energy while breaking mirror symmetry through persistent rotation, circular motion, or handed interactions \cite{Kuemmel2013, LiebchenLevis2017, LiebchenLevis2022, capriniSelfreverting2024a}. A central class of such systems is formed by particles that both self-propel and self-rotate. Their trajectories are intrinsically curved because persistent translation is coupled to a finite angular velocity, leading to circular single-particle trajectories and, at the collective level, to rotating macroclusters, microflocks, traveling bands, vortices, self-reverting vortices, and other nonequilibrium structures \cite{Kuemmel2013, LiebchenLevis2017, levisMicroflock2018, krukTraveling2020, LiebchenLevis2022, capriniSelfreverting2024a, huangDynamical2020, zhangReconfigurable2020a, sunkesularaghavendraProgrammablePersistentRandom2026, vutukuriRational2017}. In these systems, collective and phase behavior are largely governed by the competition between persistent translation and intrinsic rotation, often assisted by alignment interactions.

Active spinners form a distinct and increasingly important class of chiral active matter. In contrast to chiral self-propelled particles, these particles only rotate, but do not translate persistently. Their activity is therefore not encoded in a self-propulsion velocity but in the continuous injection of angular momentum into the surrounding medium \cite{grzybowskiDynamic2000, nguyenEmergent2014, yeoCollective2015, soniOdd2019, reevesEmergence2021, Tierno2021, Markovich2024, jiangTransient2025}. This seemingly simple change leads to a fundamentally different route to nonequilibrium organization. Experiments on colloidal and interfacial spinner systems have revealed rotating crystals and clusters, dynamic lattices, active turbulence, arrested phase separation, edge currents, and odd collective responses \cite{grzybowskiDynamic2000, Yan2015, kokotActive2017, hanReconfigurable2020, soniOdd2019, MassanaCid2021, mattichMagnetic2023, meckeSimultaneous2023, katuriControl2024a, nelsonTopological2025a}. Related experiments on living chiral crystals have demonstrated odd dynamical and mechanical responses in a biological realization \cite{Tan2022}. Complementary simulations and continuum theories have predicted hydrodynamic clustering, lanes, phase separation, active melting, edge currents, odd viscosity, and other parity-violating constitutive responses \cite{nguyenEmergent2014, yeoCollective2015, vanZuiden2016, shenHydrodynamic2020, reevesEmergence2021, Banerjee2017, hanFluctuating2021, Markovich2024, caporussoPhase2024, jiangTransient2025, Fruchart2023}. Thus, active spinners are not merely a limiting case of chiral swimmers with vanishing propulsion. Instead, they constitute a distinct and generic platform in which collective behavior is mediated by torque injection, hydrodynamic circulation, transverse forces, and non-reciprocal interactions rather than by persistent translational motion \cite{vanZuiden2016, soniOdd2019, hanReconfigurable2020, Tan2022, Markovich2024, meckeSimultaneous2023}.

While mixtures of passive particles and self-propelled active particles have been widely studied \cite{angelaniEffective2011, gokhaleDynamic2022, stenhammarActivityInduced2015, shelkeShapeAnisotropyGoverns2026,vutukuriLightswitchablePropulsionActive2020a}, 
considerably less is known about passive matter coupled to chiral self-propelled particles. Experiments, often supported by simulations, have shown that translationally active particles can mediate effective interactions, suppress or promote phase separation, and reorganize passive colloids into dynamic structures \cite{angelaniEffective2011, SchwarzLinek2012,maddenHydrodynamically2022}. Experiments with chiral swimmers have further demonstrated the formation, persistent rotation, and percolation of passive colloidal clusters \cite{kushwahaPercolation2024,groberUnconventionalColloidalAggregation2023}. Predominantly numerical and theoretical studies have predicted directed transport, odd probe responses, and coherent rotation of passive inclusions or clusters in chiral active baths \cite{aiRatchet2016, reichhardtReversibility2019,hargusOdd2025, puitandySpontaneous2025, kushwahaEmergent2026a}. In these systems, however, the active component redistributes passive matter through persistent translation, collisions, accumulation, and sustained pushing, often resulting in irregular aggregates.

By contrast, mixtures in which activity is purely rotational remain largely unexplored. Numerical hydrodynamic studies have examined inert particles in rotor suspensions and the phase-separation dynamics of attractive passive colloids driven by self-rotating particles \cite{yeoDynamics2016,yuanColloid2024}. Experimental work on spinner and passive mixtures remains scarce, with existing studies focusing either on the motion of a single passive object in a macroscopic granular spinner bath \cite{yangTopologically2021} or on the coarsening of small spinner clusters in a passive colloidal bath \cite{aragonesAggregation2019}. 
This leaves a fundamental question: can angular-momentum injection alone, without persistent self-propulsion, transfer chirality and boundary-localized chiral dynamics to ordinary passive matter? Establishing such a transfer would identify torque injection as a generic mechanism for imparting chiral functionality to material components that are not themselves driven.

Here, we combine experiments and simulations to systematically explore how passive colloidal matter organizes when mixed with magnetic active spinners. The active particles rotate in an externally imposed magnetic field but do not self-propel, whereas the passive colloids interact with each other through short-range depletion attractions. 
We demonstrate how active spinners can transfer chirality, collective rotation, and boundary currents to undriven matter. At high passive fraction and strong magnetic attraction, the active spinners self-assemble into rotating clusters that feature pronounced edge currents and are embedded in a passive matrix. Strikingly, upon changing composition and spinner attraction, this organization turns inside out: passive colloids form clusters that counter-rotate within the surrounding chiral spinner liquid.
Despite being undriven, these passive rotors exhibit edge currents and the same $|\Omega|\sim R^{-2}$ size scaling as their active counterparts, pointing to a common boundary-driven mechanism.

In this inverted regime, the passive clusters are not directly actuated by the external field. Instead, they acquire chirality through effective transverse and non-reciprocal forces exerted by the surrounding spinner bath. A minimal model based on these interactions reproduces the inside-out inversion and common boundary-driven dynamics of active and passive rotors. More broadly, our results establish torque injection as a mechanism for transmitting chiral dynamics between material components, allowing undriven matter to acquire collective rotation and boundary transport.

\section*{Chiral active rotors form within a passive matrix}

\begin{figure*}[t]
    \centering
    \includegraphics[width=\linewidth]{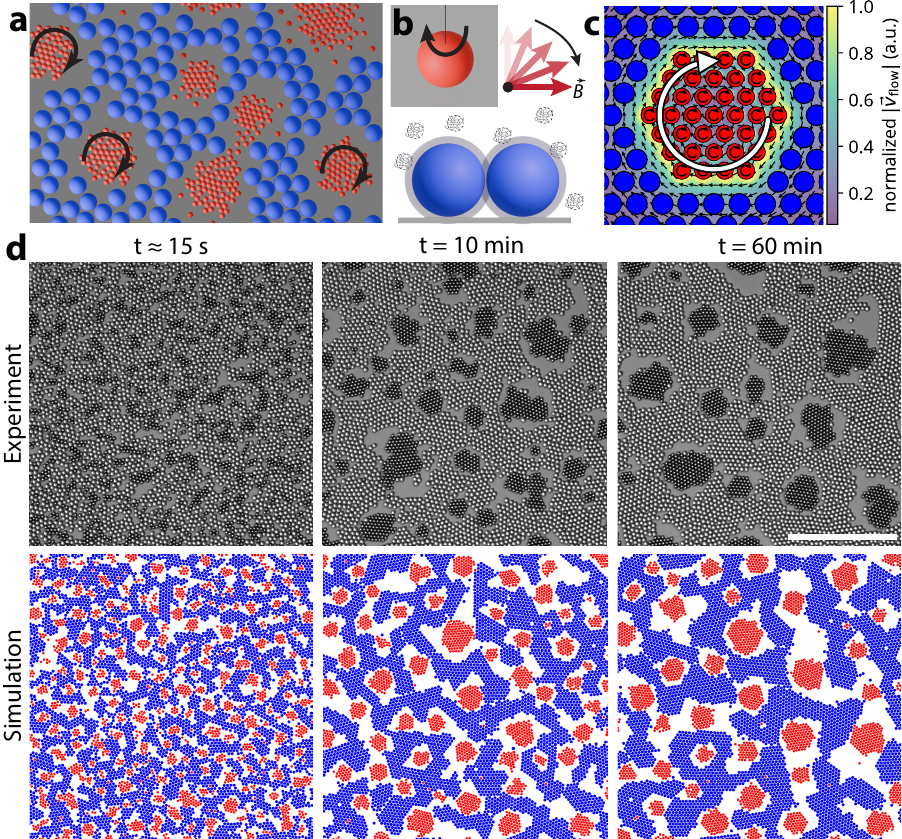}
\caption{\textbf{Chiral active rotors form and coarsen within a passive colloidal matrix.}
\textbf{a,} Schematic of the quasi-two-dimensional experimental system. Polymer-based magnetic particles (red) and passive silica colloids (blue) sediment near a glass substrate and are subjected to an in-plane magnetic field rotating clockwise (CW) when viewed from above. \textbf{b,} The magnetic particles rotate synchronously with the applied field near the substrate, injecting angular momentum into the surrounding fluid. A polymeric depletant induces short-ranged depletion attractions that act predominantly between the passive silica particles. \textbf{c,} Schematic of a chiral active rotor. The background color denotes the speed of the velocity field obtained by superposing idealized rotlet flows from all spinners. Arrows indicate the flow direction and hence the direction in which the effective transverse force acts. In the reference regime ($\phi_\mathrm{P} > 0.5$, $\beta > 1$), chiral active rotors are embedded in a passive matrix. The flows of all spinners largely cancel in the cluster interior but remain uncompensated at its outer boundary, generating a torque on the chiral active rotor in the spinner-rotation direction.  \textbf{d,} Time evolution of the mixture in experiments (top) and numerical simulations (bottom) for $\phi_\mathrm{P}=0.7$, starting from a dispersed configuration. Passive particles are shown in blue and active spinners in red in the simulations. The passive particles form a percolating, gel-like network whose voids contain CW-rotating chiral active rotors that merge and coarsen over time. Experimental times are indicated above the images. Scale bar: \SI{100}{\um}.}
    \label{fig:schematic}
\end{figure*}

We investigate binary mixtures of magnetic and silica colloids sedimented near the substrate of an observation chamber (see Methods). The translational dynamics of the resulting particle layer are effectively two-dimensional (Fig. \ref{fig:schematic}a). An in-plane magnetic field rotating clockwise (CW) drives the magnetic particles to spin synchronously without inducing self-propulsion (Fig. \ref{fig:schematic}b). We refer to the magnetic and silica particles as active spinners and passive colloids, respectively.

Polyacrylamide (PAM) serves as the polymeric depletant in the experiments, inducing short-range attraction predominantly between the passive colloids. By contrast, the active spinners are only weakly affected by depletion, which we attribute to their higher surface roughness, confirmed by scanning electron microscopy images (see SI Fig. S2). In the absence of PAM, the passive colloids do not form stable, hexagonally packed clusters (see SI Fig. S3). The active spinners also acquire induced dipole moments from the external rotating magnetic field, resulting in a time-averaged isotropic attraction. The system thus features two dominant interactions: the short-range depletion attraction between the passive particles and the long-range magnetic field-controlled attraction between the active spinners.

We characterize the composition by the proportion of the passive particles in the total area fraction, $\phi_{\mathrm P}=A_{\mathrm P}/(A_{\mathrm P}+A_{\mathrm A})$, where $A_{\mathrm P}$ and $A_{\mathrm A}$ denote the nominal projected areas covered by passive particles and active spinners, respectively. For particles of different diameters, this corresponds to $\phi_{\mathrm P} =N_{\mathrm P}d_{\mathrm P}^{2}/(N_{\mathrm P}d_{\mathrm P}^{2}+N_{\mathrm A}d_{\mathrm A}^{2})$. The total particle area fraction is denoted by $\phi=(A_{\mathrm P}+A_{\mathrm A})/A_{\mathrm{sys}}$ and fixed to $\phi\simeq0.5$ for all experiments and simulations.

For the state shown in Fig.~\ref{fig:schematic}, we used \SI{3.9}{\micro\metre} passive particles and \SI{3.3}{\micro\metre} active spinners in a $0.15$\,wt.\% polyacrylamide solution, with $\phi_{\mathrm P}\simeq0.70$. Upon applying a rotating magnetic field of order \SI{\sim 1}{\milli\tesla} at \SI{1}{\hertz}, the initially dispersed mixture develops a species-segregated morphology. The passive colloids form a dense, percolating matrix, whereas the magnetic particles assemble into finite clusters occupying the voids within this matrix (Fig.~\ref{fig:schematic}d and Supplementary Video~1). The clusters of active spinners rotate in the direction imposed by the magnetic field and coarsen over time through growth and cluster-merging events. We refer to them as chiral active rotors. This state provides the reference morphology: finite chiral active rotors embedded in a continuous passive matrix.

\section*{Minimal model for spinner-mediated chiral coupling}

To identify the minimal physical ingredients underlying the experimentally observed patterns (Fig.~\ref{fig:schematic}d), we developed an effective Brownian dynamics model for a binary mixture of passive colloids and active spinners confined to a plane. Both species are represented as overdamped, slightly soft discs. The model comprises steric repulsion between all particles, short-ranged depletion attraction strongest between passive colloids, an effective magnetic attraction between spinners, and a non-conservative transverse force generated by spinner rotation. For simplicity, the magnetic interaction between spinners is treated in a time-averaged form, justified by spinners rotating rapidly ($\tau_\mathrm{rot}\sim \SI{1}{s}$) compared with the time required to translate over one particle diameter ($\tau_\mathrm{trans}\sim \SI{30}{s}$). The resulting effective attraction is controlled by a single parameter $\beta$, which captures the strength of the magnetic coupling between spinners. 
The key nonequilibrium ingredient is the transverse force $\vec F^{\perp}$, which represents the net effect of the azimuthal flow generated by a rotating spinner. It acts perpendicular to the line connecting two particles and thereby provides a minimal description of spinner-induced chiral transport. 
Because only active spinners generate this transverse forcing, with no equal counterpart from passive particles, the active–passive coupling is intrinsically non-reciprocal. Its strength is controlled by an effective coupling parameter $\lambda$, which subsumes hydrodynamic, near-wall, and substrate-dependent details. Thus, the conservative interactions determine which species clusters, whereas the transverse non-reciprocal forcing endows the resulting structures with chiral dynamics.
Despite its minimal character, the model reproduces the experimentally observed formation and coarsening of rotating active clusters embedded in a passive matrix (Fig.~\ref{fig:schematic}d, Supplementary Video~2). This indicates that passive depletion attraction, spinner attraction, thermal fluctuations, and spinner-induced transverse forcing are sufficient to account for the observed chiral organization, without explicitly resolving the full hydrodynamic many-body interactions of the experimental system. We next use this model to explore how the identity and dynamics of the rotating phase change when composition and spinner attraction are varied.

\section*{Active spinners drive passive matter into chiral rotors}

\begin{figure}[t]
    \centering
    \includegraphics[width=\linewidth]{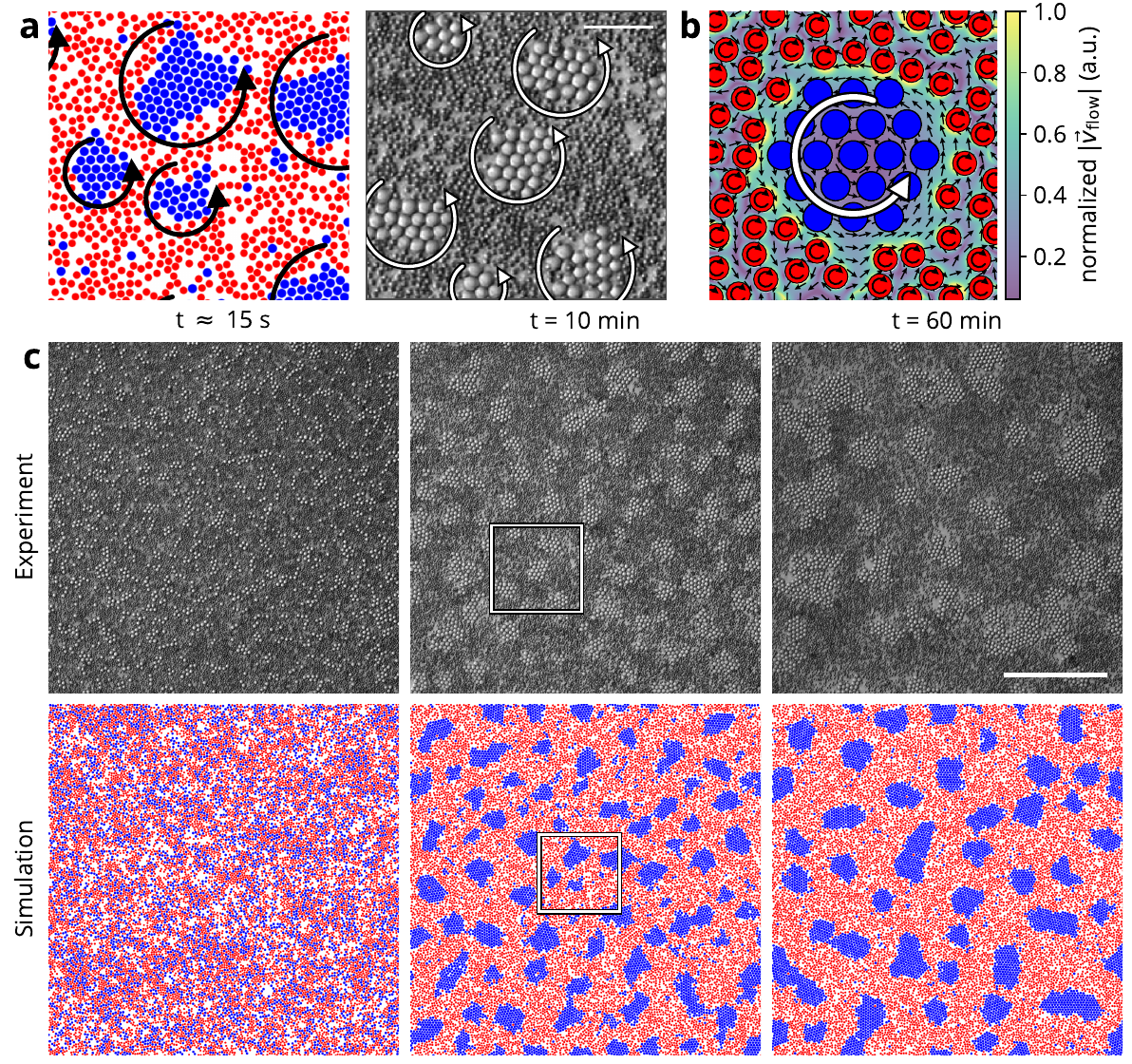}
    \caption{\textbf{Emergence of passive chiral rotors in an active-spinner fluid}.
    \textbf{a,} Insets for an experimental micrograph and a simulation snapshot at $\phi_\mathrm{P}=0.3$, with $\beta=0.5$ and $\alpha_\mathrm{pp}=3.7$ in the simulation, showing passive particles forming finite clusters rotating CCW embedded in a dense background of spinners rotating CW.  Scale bar: \SI{20}{\um}. \textbf{b,} Schematic of a passive rotor. The background color denotes the speed of the velocity field obtained by superposing idealized rotlet flows from all spinners. Arrows indicate the flow direction and hence the direction in which the effective transverse force acts. In the inverted regime ($\phi_\mathrm{P}<0.5$, $\beta < 1$), passive rotors can be viewed as inclusions within a chiral spinner liquid. The uncompensated flow at the inner boundary of the spinner liquid exerts a torque opposite to the spinner's rotation direction, causing the passive rotors to rotate CCW.
    The lower flow magnitude around passive rotors indicates a smaller transverse force per unit boundary length than for chiral active rotors, consistent with their lower angular velocity (see Fig. \ref{fig:cluster-rotation}). \textbf{c,} Time evolution of the mixture in experiments (top) and numerical simulations (bottom) for $\phi_\mathrm{P}=0.3$, starting from a dispersed configuration. Passive particles are shown in blue and active spinners in red in the simulations. The passive particles form clusters rotating CCW submerged in a fluid of active spinners. Experimental times are indicated above the images. Scale bar: \SI{100}{\um}.}
    \label{fig:inverted}
\end{figure}

A striking prediction of the model is an inside-out inversion of the rotating phase, i.e., that the identity of the rotating clustered phase can be reversed. At low passive-particle fraction and weak magnetic attraction between spinners ($\phi_\mathrm{P}=0.3$, $\beta=0.5$), the spinners no longer condense into compact active clusters. Instead, they form a connected chiral spinner liquid, while the passive colloids assemble into finite clusters that rotate coherently in the opposite direction (see Fig.~\ref {fig:inverted}a, Supplementary Video~4). Thus, rather than CW-rotating chiral active rotors embedded in a passive matrix, the system forms CCW-rotating passive rotors embedded within an active-spinner fluid.

To realize this regime experimentally, we reduce the effective magnetic attraction by using smaller magnetic particles with a diameter of \SI{1.2}{\micro\metre}. Simply lowering the field strength for the larger particles is not sufficient, because synchronous rotation with the external field is lost before the desired weak-attraction regime is reached. Under these modified conditions, the experiments reproduce the predicted inside-out morphology (Fig.~\ref{fig:inverted}c). The passive clusters have no direct magnetic coupling to the external field, yet rotate coherently opposite to the surrounding spinners: while the field and the spinners rotate clockwise, the passive rotors rotate counterclockwise (see Supplementary Video~3). The inversion is therefore both structural and dynamical, with the undriven passive component becoming the coherently rotating phase.

In the model, at weak spinner attraction, thermal fluctuations prevent the magnetic particles from forming compact clusters, allowing them to remain as the continuous phase. The passive particles, in contrast, still experience depletion attraction and therefore assemble locally. Because their area fraction is low, they do not form a system-spanning gel but instead form finite clusters. The active- and passive-rotor states can be understood as complementary boundary geometries. A chiral active rotor presents an outer boundary of the spinner phase, whereas a passive rotor forms an inclusion--an inner boundary--within the spinner liquid. In both cases, spinner-induced transverse forces largely cancel in the bulk but remain uncompensated at the boundary. Accordingly, at the outer boundary of an active rotor, the resulting rim torque acts in the direction of spinner rotation (Fig.~\ref{fig:schematic}c) \cite{Yan2015}. Turning this geometry inside out reverses the boundary orientation, and hence the torque: the surrounding spinner liquid drives the passive rotor in the opposite direction (Fig.~\ref{fig:inverted}b). In the passive-rotor state, the density of torque-generating spinners at the interface is lower than at the rim of a compact chiral active rotor. Consequently, the resulting interfacial torque density, and hence the rotation rate, is correspondingly smaller. However, because the driving remains localized at the boundary in both morphologies, we expect active and passive rotors to exhibit the same dependence of angular velocity on cluster size, differing primarily in the prefactor.


\begin{figure}[htbp]
    \centering
    \includegraphics[width=0.6\textwidth]{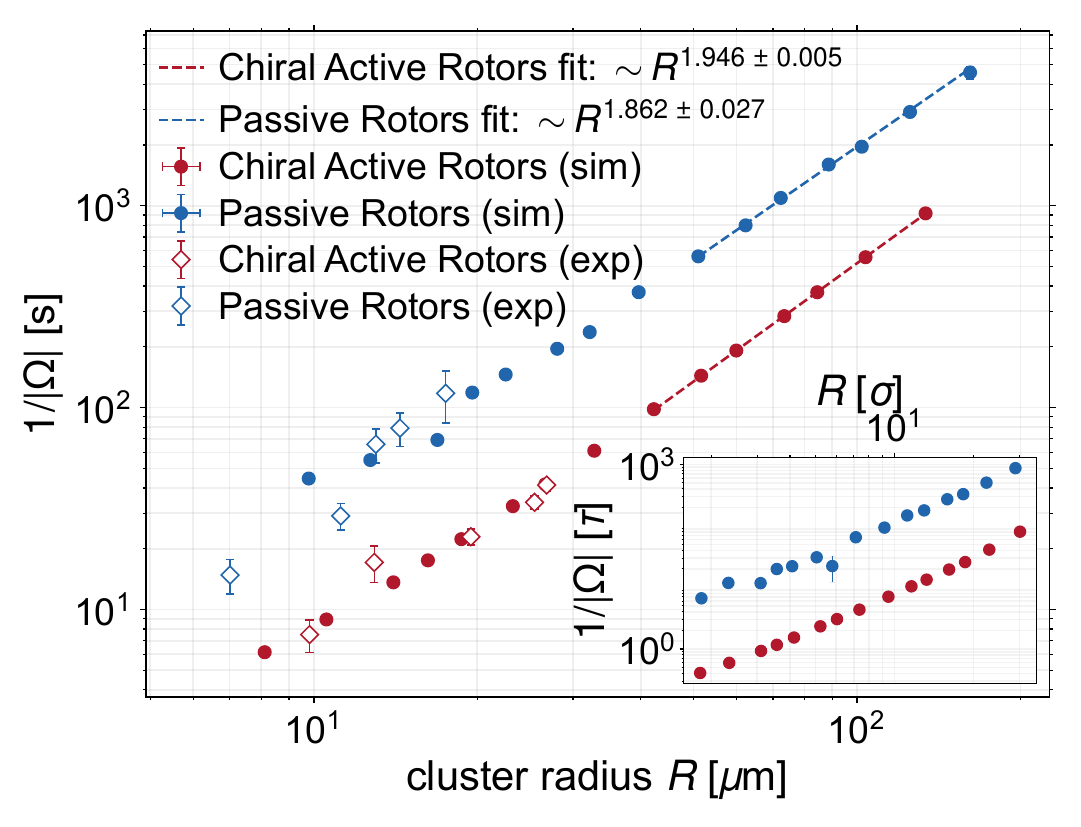}
    \caption{\textbf{Rotational timescale versus cluster size for active and passive rotors.}
    The rotational timescale $1/|\Omega|$ increases approximately quadratically with cluster radius for both chiral active and passive rotors. Filled circles show simulations and open diamonds show experiments; red and blue denote active and passive rotors, respectively. Simulation error bars are sample standard errors across five independent runs, whereas experimental vertical error bars are obtained from the uncertainty in $\Omega$. Dashed lines show weighted fits $T\sim R^n$ to the seven largest-radius simulation points, yielding $n=1.946\pm0.005$ for active rotors and $n=1.862\pm0.027$ for passive rotors. Passive rotors rotate more slowly than active rotors of comparable size, consistent with a smaller effective driving torque. The transverse-interaction strength was calibrated separately for active and passive rotors; see Methods for parameter choices and conversion to physical units. Inset: simulations varying only $\phi_\mathrm{P}$ and $\beta$, without independent calibration, qualitatively reproduce the slower passive-rotor dynamics. Error bars smaller than the markers are not visible.}
    \label{fig:cluster-rotation}
\end{figure}

To test this common boundary-driven picture, we measure the magnitude of the mean angular velocity $|\Omega|$ as a function of cluster radius $R$ in both experiments and simulations. Figure~\ref{fig:cluster-rotation} shows the corresponding rotational timescale $1/|\Omega|$ for chiral active and passive rotors. For both types of rotor, $1/|\Omega|$ increases approximately linearly with $R^2$, corresponding to the scaling $|\Omega|\propto R^{-2}$.

For chiral active rotors, this scaling is consistent with the rim-torque scaling reported in \cite{Yan2015}. Transverse forces approximately cancel in the bulk but remain uncompensated at the boundary (Fig.~\ref{fig:schematic}c). If the tangential force per boundary particle is of order $F^\perp$, the number of torque-generating particles scales as $n_{\rm edge}\propto R$. Since these tangential boundary forces act at a radial distance of order $R$ from the center of rotation, the resulting driving torque scales as $\tau^\perp \sim n_{\rm edge} F^\perp R \sim F^\perp R^2$.

The rotational drag, by contrast, is generated throughout the cluster. A particle at distance $r$ from the cluster center moves with tangential velocity $v_\theta=\Omega r$ and contributes a drag torque of order $\gamma \Omega r^2$. Integration over the cluster area gives $\tau_{\rm drag}\sim \gamma \Omega R^4$. Balancing driving and drag therefore yields $\Omega \sim R^{-2}$, or equivalently $1/\Omega \propto R^2$, consistent with the dependence observed in Fig.~\ref{fig:cluster-rotation}.

The same geometric argument can be applied to passive rotors. Their driving torque is generated by transverse forces exerted by the surrounding spinner fluid on the passive-cluster boundary (Fig.~\ref{fig:inverted}b), whereas their rotational drag remains distributed throughout the cluster. Active and passive rotors, therefore, share the same size-scaling exponent. At a given cluster radius, however, passive rotors exhibit systematically longer rotational times, corresponding to a larger scaling prefactor and hence weaker effective interfacial driving.

The common $R^{-2}$ scaling is consistent with a boundary-localized torque as the mechanism controlling the mean rotation of both rotor types. However, it does not reveal how the resulting motion is distributed within the clusters. We therefore resolve the particle dynamics within a cluster to determine whether the presumed interfacial driving also produces edge currents, i.e., boundary-localized transport beyond rigid-body rotation.

\section*{Passive rotors exhibit edge currents}

Edge currents are a characteristic feature of active-spinner materials, where the rotational activity gives rise to directed motion along boundaries \cite{vanZuiden2016, soniOdd2019, katuriControl2024a, nelsonTopological2025a, MassanaCid2021, caporussoPhase2024}. 
In a rotating cluster, such currents appear as an excess tangential velocity beyond the rigid-body contribution $v_\theta = \Omega r$. We now ask whether the undriven passive rotors inherit this characteristic boundary dynamics from the spinners. 

\begin{figure}[htbp]
    \centering
    \includegraphics[width=\linewidth]{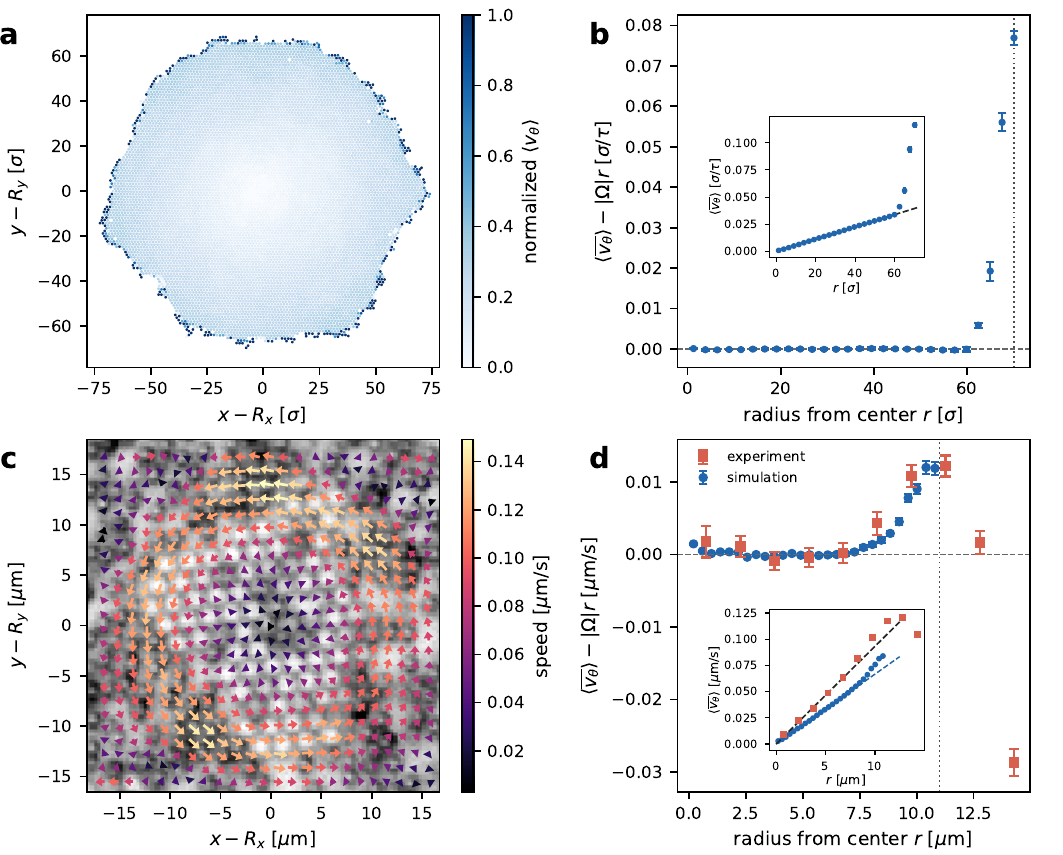}
    \caption{\textbf{Edge currents in passive rotors.}
    \textbf{a,} Simulated passive rotor at reduced passive-particle attraction, $\alpha=2.8$. Particles are colored by their time-averaged tangential velocity $v_\theta$, normalized by the maximum of the radially averaged profile.
    \textbf{b,} Azimuthally averaged excess tangential velocity $\langle v_\theta\rangle-|\Omega|r$, obtained by subtracting the solid-body rotation of the cluster. The pronounced increase near the rim reveals a boundary-localized edge current. The inset shows the corresponding tangential velocity $\langle v_\theta\rangle$; the dashed line denotes the solid-body contribution $|\Omega|r$. 
    \textbf{c,} Bright-field micrograph of an experimental passive rotor overlaid with the time-averaged PIV velocity field. Arrow directions indicate the local flow, and colors denote the velocity magnitude.
    \textbf{d,} Azimuthally averaged excess tangential velocity $\langle v_\theta\rangle-|\Omega|r$ of the experimental rotor, revealing enhanced tangential motion near the cluster boundary. The inset shows the corresponding tangential velocity $\langle v_\theta\rangle$, with the dashed line indicating the solid-body contribution $|\Omega|r$, where $|\Omega|=\SI{0.0093}{rad.s^{-1}}$. Simulation data for a cluster of similar size underestimates the absolute azimuthal velocity but agrees well with the excess tangential velocity. Error bars denote the standard uncertainty of the mean. The dotted vertical line indicates the cluster radius.}
    \label{fig:edge-currents}
\end{figure}

To resolve the edge current clearly in simulations, we reduce the passive-particle attraction to $\alpha=2.8$ for a cluster of approximately $N=10,000$ particles (see Supplementary Video~6). This reduction enhances particle rearrangements at the boundary while preserving a cohesive passive rotor.
The time-averaged tangential velocity (Fig.~\ref{fig:edge-currents}a) reveals pronounced edge currents. In the interior, its radial profile closely follows the linear increase expected for a solid-body rotation. Near the rim, however, the velocity rises sharply above $\lvert\Omega\rvert r$, revealing a pronounced edge current (Fig.~\ref{fig:edge-currents}b). 

To experimentally resolve boundary motion in individual passive rotors, we required depletion attractions that were sufficiently strong to stabilize a rigid, hexagonally packed core, yet sufficiently weak that interfacial driving induces particle hopping at the boundaries (see Supplementary Video~5). We achieved this balance using silica particles of diameter $\SI{2.1}{\um}$, a PAM concentration of $0.18\,\mathrm{wt}\,\%$, and a magnetic-field rotation frequency of $\SI{1.3}{Hz}$. The time-averaged velocity field shows coherent azimuthal motion with enhanced speeds around the cluster periphery (Fig.~\ref{fig:edge-currents}c). Correspondingly, the tangential velocity follows the solid-body prediction in the cluster interior but exceeds it as the boundary is approached (orange data points in Fig.~\ref{fig:edge-currents}d). Because these experimentally accessible clusters are relatively small ($N \approx 80$), we compare them to a simulation of a comparably sized passive rotor, which shows a similar boundary-localized excess-velocity profile after rescaling lengths to the experimental cluster radius (see Methods).

Passive rotors, therefore, do not merely rotate coherently: despite being undriven, they inherit the boundary-localized current characteristic of active-spinner clusters. The inside-out state thus transfers not only chirality and collective rotation, but also a spatially organized mode of boundary transport from the active spinner phase to passive matter.

\newpage

\section*{State diagram of active-passive organization}

\begin{figure}[htbp]
    \centering
    \includegraphics[width=\linewidth]{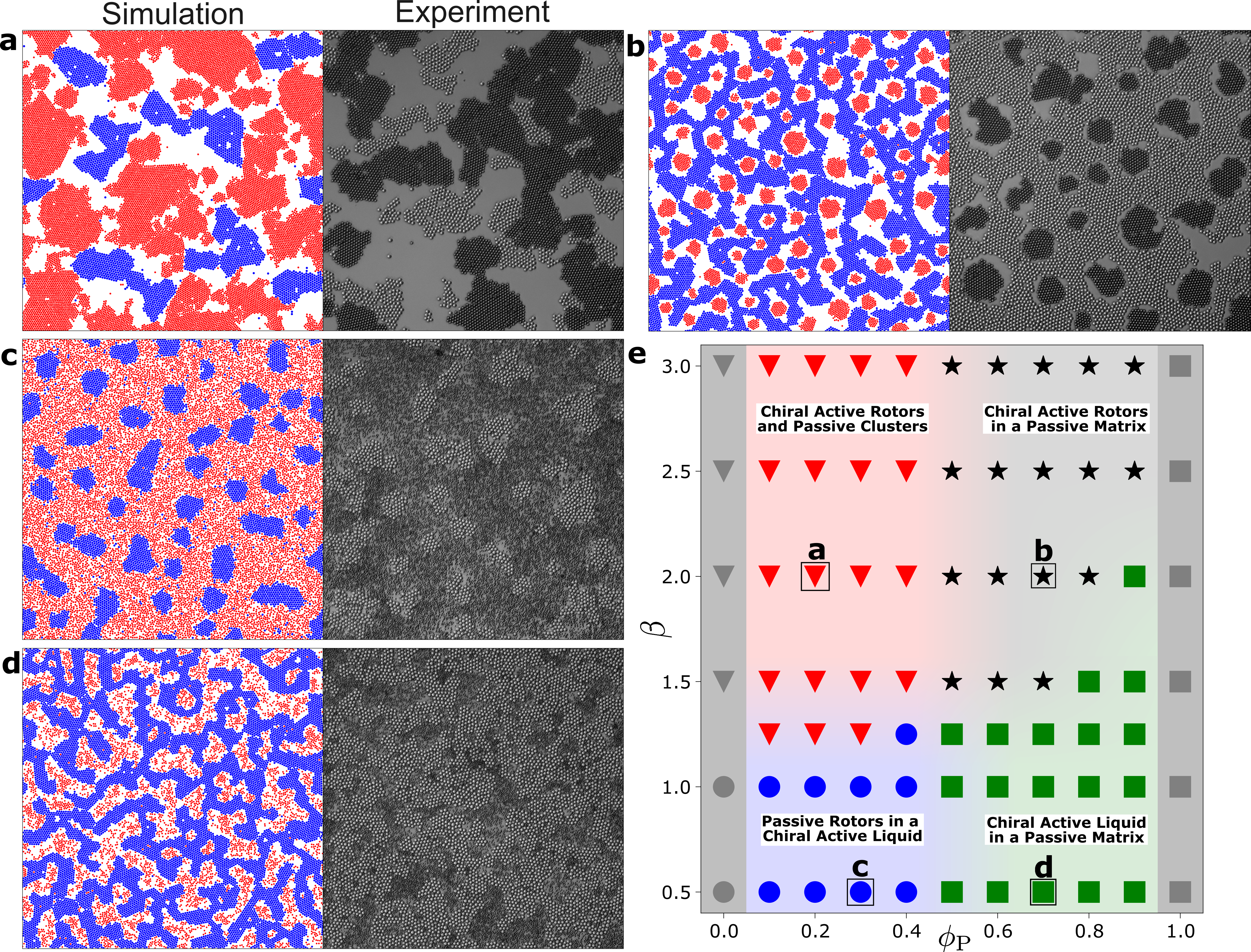}
    \caption{\textbf{State diagram of active spinner and passive colloid mixtures.}
    \textbf{a,} At low $\phi_{\mathrm{P}}$ and strong magnetic attraction, active and passive particles form coexisting clusters, with chiral active rotors rotating CW intermediately and becoming mostly arrested. Passive clusters form in regions depleted by the spinners. \textbf{b,} At high $\phi_{\mathrm{P}}$ and strong attraction, finite CW-rotating chiral active rotors are embedded in a passive matrix. \textbf{c,} At low $\phi_{\mathrm{P}}$ and weak attraction, finite CCW-rotating passive rotors are embedded in a connected chiral spinner liquid. \textbf{d,} At high $\phi_{\mathrm{P}}$ and weak attraction, a chiral spinner liquid permeates the voids of a passive matrix. 
    \textbf{e,} Collective states observed in simulations as a function of the passive-particle area fraction $\phi_{\mathrm{P}}$ and non-dimensional magnetic attraction strength $\beta$ at fixed total area fraction $\phi=0.5$. Symbols mark representative state points; corresponding simulation snapshots and experimental micrographs are arranged around the diagram. Assignment of data points to the respective regions is based on percolation and cluster analysis (see SI). Grey regions and symbols denote the corresponding single-species states (see SI Fig. S8). The experiments are realizations of representative states without a quantitative mapping of $\beta$.}
    \label{fig:diagram}
\end{figure}

To explore how the inside-out inversion fits into the broader collective behavior of the mixture, we systematically vary both the area fraction of passive particles $\phi_\mathrm{P}$ and the strength of the effective magnetic attraction between spinners, $\beta$, in simulations. The resulting $\beta$ vs. $\phi_\mathrm{P}$ state diagram is shown in Fig. \ref{fig:diagram}. We experimentally realize representative states by varying composition, magnetic field strength, and magnetic-particle size, without attempting a quantitative mapping onto $\beta$. 

To obtain states c and d experimentally, we use the \SI{1.2}{\micro\metre} magnetic particles, as these particles do not cluster but still rotate synchronously with the applied magnetic field. In contrast, simulations allow for more systematic variation of $\beta$ and $\phi_\mathrm{P}$ and a more in-depth analysis, without the limitation of having to change the particle size. Beyond the two regimes already discussed (Fig. \ref{fig:diagram}b,c), two further regimes are obtained when tuning the fraction of passive particles or the magnetic attraction strength. At low fraction of passive particles and sufficiently strong magnetic attraction, both species form clustered structures, leading to a mixed clustered regime in which passive and active clusters coexist (Fig. \ref{fig:diagram}a, Supplementary Videos 7,8). At high passive fraction but weak magnetic attraction, the spinners no longer condense into finite chiral active rotors. Instead, they remain less well-structured and occupy the voids and channels of the passive matrix, which itself remains connected due to depletion-induced attraction. In this regime, the passive phase still provides the structural scaffold, while the active component forms a chiral fluid moving through it (Fig. \ref{fig:diagram}d, Supplementary Videos 7,8). The state diagram, therefore, shows that the system is not organized simply by whether one or the other species clusters. Rather, changes in composition and attraction strength can switch which species form the clustered rotating phase and which form the continuous scaffold. The state diagram thus shows that the inside-out inversion is part of a broader switch in the roles of the two components: composition and spinner attraction determine whether active or passive matter forms the continuous phase and which component organizes into rotating inclusions.

\section*{Conclusions}
We have shown that purely rotational activity can transfer chirality, collective rotation, and boundary currents to undriven colloidal matter. Upon changing composition and spinner attraction, the organization of the mixture turns inside out: rotating active clusters embedded in a passive matrix are replaced by passive rotors immersed in a chiral spinner liquid. Despite being undriven, these passive rotors exhibit edge currents and the same $|\Omega| \sim R^{-2}$ size scaling as their active counterparts, pointing to a common boundary-driven mechanism.

A minimal model shows that spinner-mediated transverse and non-reciprocal interactions are sufficient to reproduce these complementary chiral states. Active and passive rotors can be viewed as outer and inner boundaries of the spinner phase, respectively, such that the same transverse forcing generates torques of opposite sign. This establishes angular-momentum injection as a mechanism for transmitting chiral dynamics between material components without persistent self-propulsion or direct actuation of the component carrying the resulting motion.

More broadly, our results reveal a design principle for composite nonequilibrium materials: the active component that injects energy and angular momentum need not itself form the structure that carries the resulting mechanical functionality. Instead, an active environment can endow ordinary matter with collective motion and boundary transport through emergent interactions at their interface. This separation between activity injection and functional response opens possibilities for controlling chiral transport in materials whose structure, interactions, and mechanical properties can be chosen independently of the active component. Future work could exploit this principle to explore odd mechanical responses, different inclusion shapes and rigidities, and the role of confinement and curvature in controlling boundary-driven transport.


\newpage

\section*{Materials and Methods}

\subsection*{Materials}

Two sizes of carboxyl-functionalized magnetic particles were used: \SI{1.2}{\micro\metre} polystyrene-based particles with a magnetite core (Microparticles GmbH) and \SI{3.3}{\um} ProMag\textsuperscript{\textregistered} HP particles (Polysciences, Inc.). Silica particles of sizes \SI{3.9}{\um} and \SI{2.1}{\um} were also obtained from Microparticles GmbH. A 10 wt.-\% aqueous solution of polyacrylamide (molecular weight: \qtyrange{700000}{1000000}{\g/\mol}) was purchased from Polysciences, Inc. All aqueous solutions were prepared using ultrapure Milli-Q\textsuperscript{\textregistered} water, and all chemicals were used as received unless otherwise stated.

\subsection*{Sample preparation}
Experiments were performed in custom observation chambers fabricated by gluing the base of a pipette tip to a cover glass. Before fabrication, the coverslips were cleaned for \SI{20}{min} using a plasma cleaner (Bioforce Nanosciences, UV/Ozone ProCleaner Plus). The wide end of a \SI{200}{\micro\litre} pipette tip was cut off and glued to the cleaned coverslip using a UV-curable adhesive (Norland Optical Adhesive 68) by curing it under a \SI{365}{\nm} UV LED for 5 minutes.

To this well, \SI{40}{\micro\litre} of magnetic particles in 0.15 wt.-\% polyacrylamide (PAM) were added and left to sediment for up to \SI{3}{\hour}, depending on particle size. During this time, the well was sealed using Parafilm\textsuperscript{\textregistered}. Then, \SI{40}{\micro\litre} of silica particles in 0.15 wt.-\% PAM was added gently, to not disturb the layer of magnetic particles at the bottom of the well. The silica particles were allowed to settle over the course of \SI{10}{\min}, after which the experiment was started. To prevent evaporation during the experiment, the chamber was again sealed with Parafilm\textsuperscript{\textregistered}. For the edge current experiments, 0.18 wt.-\% PAM was used.

\subsection*{Magnetic field setup}
A rotating magnetic field at constant angular velocity was applied to induce spinning of the magnetic particles \cite{sunkesularaghavendraProgrammablePersistentRandom2026}. The field was generated by a permanent magnet mounted on the rotor of a servo motor (Parallel Feedback 360\textdegree~High-Speed Servo), positioned several centimeters above the sample. The magnet's north-south axis was oriented parallel to the substrate and perpendicular to the motor's rotation axis. The rotation speed of the magnetic field was controlled by operating the servo via an Arduino Uno R3 microcontroller equipped with a PWM shield (Adafruit 16-Channel PWM/Servo Shield). The magnetic field strength was varied by tuning the distance between the magnet and the coverslip.

\subsection*{Microscopy imaging and particle tracking}
The particles were imaged using a Nikon Eclipse TE2000-U inverted fluorescence microscope equipped with either a 20x (N.A. 0.5) or a 40x (N.A. 0.75) air objective lens and a Basler acA4112–30um CMOS camera. The objective was centered on the middle of the chamber to minimize wall effects. Time-lapse image sequences were recorded in bright-field mode at 1 frame per second (fps). The particles were tracked from the image sequences using the Difference of Gaussians (DoG) detector in the TrackMate plugin in Fiji (ImageJ) \cite{schindelinFiji2012,tinevezTrackMate2017}.

\subsection*{Effective particle-based model}

We model the magnetic and passive colloids as two species of overdamped discs with diameters $\sigma_{\mathrm a}$ and $\sigma_{\mathrm p}$, respectively. The position of particle $i$ evolves according to 
\begin{equation}
    \gamma \dot{\vec r}_i = -\sum_{j\neq i}\nabla_i U_{s_i s_j}(r_{ij}) + \sum_{\substack{j\in\mathrm a\\j\neq i}} \vec F^\perp(\vec r_{ij}) + \sqrt{2\gamma k_{\mathrm B}T}\, \vec{\xi}_i(t), 
\label{eq:eom} 
\end{equation} 
where $s_i\in\{\mathrm a,\mathrm p\}$ denotes the particle species, $\vec r_{ij}=\vec r_i-\vec r_j$, and $\vec{\xi}_i$ is Gaussian white noise satisfying $\langle \vec{\xi}_i(t)\rangle=0$ and $\langle \xi_{i\mu}(t)\xi_{j\nu}(t')\rangle =\delta_{ij}\delta_{\mu\nu}\delta(t-t')$. The conservative pair potential is 
\begin{equation} 
    U_{ss'}(r) = U^{\mathrm{WCA}}_{ss'}(r) + U^{\mathrm{depl}}_{ss'}(r) + \delta_{s\mathrm a}\delta_{s'\mathrm a} U^{\mathrm{dip}}(r). 
\label{eq:pairpotential} 
\end{equation} 
Excluded-volume interactions are described by the Weeks--Chandler--Andersen potential 
\begin{equation} U^{\mathrm{WCA}}_{ss'}(r) = 
    \begin{cases} 
        4\varepsilon \left[ \left(\dfrac{\sigma_{ss'}}{r}\right)^{12} - \left(\dfrac{\sigma_{ss'}}{r}\right)^6 \right] +\varepsilon, & r<2^{1/6}\sigma_{ss'},\\[4pt] 0, & r\geq2^{1/6}\sigma_{ss'}, 
    \end{cases} 
\end{equation} 
with $\sigma_{ss'}=(\sigma_s+\sigma_{s'})/2$. The depletion-induced attraction between particles is represented by \begin{equation} 
    U^{\mathrm{depl}}(r) = 
    \begin{cases} 
        -\alpha_{ss'} \tanh\!\left[\Delta(r^{\mathrm d}_{ss'}-r)\right], & r\leq r^{\mathrm d}_{ss'},\\[4pt] 0, & r>r^{\mathrm d}_{ss'}, 
    \end{cases} 
\label{eq:depletion} 
\end{equation}
where $\alpha_{ss'} $, $r^{\mathrm d}_{ss'}$, and $\Delta$ set the attraction strength, range, and steepness, respectively. Depletion attractions act between all species, but interactions involving active spinners are chosen to be of order $k_\mathrm{B}T$, such that depletion predominantly controls the cohesion of passive particles. Rapid rotation of the magnetic particles allows their time-dependent dipole--dipole interaction to be replaced by its rotational average (see SI), yielding 
\begin{equation} 
    U^{\mathrm{dip}}(r) = -\beta k_{\mathrm B}T \left(\frac{\sigma_{\mathrm a}}{r}\right)^3 , \label{eq:dipolar} 
\end{equation} 
where $\beta$ controls the effective magnetic attraction. The flow generated by each active spinner is represented by an effective transverse interaction 
\begin{equation} 
    \vec F^\perp(\vec r_{ij}) = \lambda\frac{k_{\mathrm B}T}{\sigma_{\mathrm a}} \left(\frac{\sigma_{\mathrm a}}{r_{ij}}\right)^2 \hat{\vec z}\times\hat{\vec r}_{ij}, \qquad j\in\mathrm a , \label{eq:transverse_force} 
\end{equation} 
where $\lambda$ controls the coupling strength. Active particles therefore exert transverse forces on both species, whereas passive particles generate no corresponding interaction. The active--passive transverse coupling is consequently non-reciprocal. 
The transverse interaction is motivated by the Stokes flow generated by a rotating sphere, whose leading free-space rotlet flow is azimuthal and decays as $u_\theta\propto r^{-2}$ \cite{KimKarrila1991}. Near a no-slip substrate, the corresponding image system modifies this spatial dependence \cite{Blake1971}; Eq.~\eqref{eq:transverse_force} therefore retains only the minimal rotlet-like structure rather than representing the full near-wall hydrodynamic interaction.
Lengths, energies and times are expressed in units of $\sigma_{\mathrm a}$, $k_{\mathrm B}T$ and 
\begin{equation} 
    \tau_{\mathrm B} = \frac{\gamma\sigma_{\mathrm a}^2}{k_{\mathrm B}T}, 
\end{equation} 
respectively. We henceforth set $\sigma_{\mathrm a}=k_{\mathrm B}T=\gamma=1$.

\subsection*{Brownian dynamics simulations}

Brownian dynamics simulations were performed using LAMMPS \cite{LAMMPS} with a custom pair style implementing the conservative and transverse interactions defined above. The particles were confined to a two-dimensional square domain with periodic boundary conditions. Because the transverse interaction is non-reciprocal, Newton's third-law optimization was disabled. The equations of motion were integrated with a time step $\Delta t=10^{-4}\tau_{\mathrm B}$ using the LAMMPS \texttt{brownian/sphere} integrator, with $k_{\mathrm B}T=\gamma=1$.

Unless stated otherwise, we used $\sigma_{\mathrm a}=1$, $\sigma_{\mathrm p}=1.2$, $\varepsilon=10$, and a transverse-force amplitude $\lambda=-2$. The depletion parameters were $\alpha_{\mathrm{aa}}=\alpha_{\mathrm{ap}}=1$, $\Delta=5$, and $r^{\mathrm d}_{\mathrm{aa}}=1.3$, $r^{\mathrm d}_{\mathrm{ap}}=1.4$, $r^{\mathrm d}_{\mathrm{pp}}=1.5$.

The active--active transverse interaction was truncated at $r=3$, whereas all other interactions were truncated at $r=1.5$. The passive--passive attraction $\alpha_{\mathrm{pp}}$ and the magnetic attraction $\beta$ were varied as indicated in the corresponding figures.

For the morphology simulations, $N=16{,}600$ particles were initially distributed randomly at a total covered-area fraction $\phi=0.49$. Particle overlaps were removed by energy minimization, followed by a WCA-only Brownian equilibration of $10^{5}$ time steps. The remaining interactions were then activated, and the system was evolved for up to $10^{8}$ time steps. The inverted state shown in Fig.~\ref{fig:inverted} was obtained for $\phi_{\mathrm P}=0.3$, $\beta=0.5$, and $\alpha_{\mathrm{pp}}=3.7$.

\subsection*{Cluster rotation analysis}

\paragraph{Experiments}
To estimate the angular velocity ($\Omega$), the instantaneous velocities ($v$) of tracks corresponding to particles in an isolated cluster were extracted. The center of mass of this cluster in each frame was estimated and was used to calculate the radial distance ($r$). The tangential components of the instantaneous velocities ($v_\theta$) were grouped into discrete radial bins of size $5 \mu\mathrm{m}$. The mean tangential velocity and its standard deviation were calculated for each bin. A weighted least-squares regression was then performed on these binned averages to obtain $\Omega$ for clusters of different sizes, using the standard deviations as statistical weights. The cluster radius was estimated from the particle count as $R=(d/2)\sqrt{N/\phi}$, using $d=3.3\,\mu\mathrm m$ for active spinners and $d=3.9\,\mu\mathrm m$ for passive particles, and $\phi$ the local packing fraction measured directly from the corresponding simulated clusters ($\phi=0.72$ for active rotors, $\phi=0.70$ for passive rotors; see SI). Rotational timescales were calculated as $T=1/|\Omega|$, with uncertainty $\delta T=\delta\Omega/|\Omega|^2$ where $\delta \Omega$ is the standard error of the weighted fit.

\paragraph{Simulations}

For the main-panel simulations, chiral active-rotor clusters were studied at passive-particle fraction $\phi_\mathrm{P}=0.7$, magnetic coupling $\beta=6.0$, and driving strength $\lambda=-17.4$, with tracked-particle diameter $D_\mathrm{c}=\sigma$. Passive-rotor clusters were studied at $\phi_\mathrm{P}=0.2$, $\beta=0.5$, and $\lambda=-2.5$, with $D_\mathrm{c}=3.25\sigma$, matching the experimental silica-to-spinner size ratio. The passive driving strength was chosen by matching the simulated rotational timescale to experiment. Five statistically independent Brownian-noise realizations were simulated for each cluster size.

Compact clusters of $N=15$--$5000$ tracked particles were initialized on a hexagonal lattice within a random bath at total area fraction $\phi=0.5$, energy-minimized, and relaxed for $10^5$ Brownian-dynamics steps without driving. Production trajectories spanned approximately $200$--$922\tau$ for active clusters and $2000$--$10\,000\tau$ for passive clusters, using time steps $2\times10^{-5}\tau$ and $5\times10^{-5}\tau$, respectively. Segments containing job-continuation discontinuities were excluded; passive $N=1500$, $2000$, and $3000$ trajectories were therefore restricted to steps $0$--$170\times10^6$, $0$--$130\times10^6$, and $0$--$88\times10^6$.

At each stored frame, the largest connected component of the tracked species was identified using a neighbor cutoff $1.25D_\mathrm{c}$. Coordinates were unwrapped across periodic boundaries and centered on the instantaneous cluster centroid. The rigid-body rotation between consecutive frames was obtained from the exact two-dimensional best-fit rotation, $\Delta\theta_k=\operatorname{atan2}[\sum_i(\vec r_{i,k}\times\vec r_{i,k+1})_z,\sum_i\vec r_{i,k}\cdot\vec r_{i,k+1}]$, using particles present in both frames. The accumulated angle $\theta(t)=\sum_k\Delta\theta_k$ was unwrapped, and $\Omega$ was obtained from its least-squares slope. This global estimator avoids the singular $1/r_i^2$ weighting of particle-wise angular-velocity estimators. Cluster size was quantified by $R_g^2=\langle|\vec r_i|^2\rangle/D_\mathrm{c}^2$.

The analysis window was selected independently for each realization from candidate suffixes beginning at fixed trajectory deciles. The earliest suffix was retained if its first- and second-half angular velocities differed by at most $20\%$, the fractional linear drift of $R_g^2$ was at most $2\%$, and its angular velocity agreed with that of the terminal $20\%$ within $20\%$; otherwise the terminal $20\%$ was used. We additionally verified consistency across lags of 1, 5, 10, 25, and 50 stored frames, absence of angular aliasing, and resolution of at least one net rotation. All retained trajectories passed these checks except passive $N=3000$, for which the restart-safe segments contained $0.84$--$0.86$ rotations but satisfied all stationarity criteria and gave a relative seed-to-seed SEM of $0.44\%$.

Reduced units were converted using the Stokes--Einstein time $\tau_\mathrm{phys}=3\pi\eta\sigma^3/(k_\mathrm{B}T)$ with $\eta=8.9\times10^{-4}\,\mathrm{Pa\,s}$ and $T=298\,\mathrm K$. We used $\sigma=3.28\,\mu\mathrm m$ for active clusters and $\sigma=1.20\,\mu\mathrm m$ for passive clusters, such that $3.25\sigma=3.90\,\mu\mathrm m$. Each realization was converted separately using $R_i=\sqrt{2}\,D_\mathrm{c}^{*}\sigma\sqrt{R_{g,i}^2}$ and $T_i=\tau_\mathrm{phys}/|\Omega_i|$, where $D_\mathrm{c}^{*}=D_\mathrm{c}/\sigma$ and $\sqrt{2}$ converts the radius of gyration to the outer radius of a uniformly filled disk. Reported values and error bars are the mean and sample standard error across the five realizations.

The seven largest-radius points of each species were fitted to $T\sim R^n$ by weighted least squares in logarithmic coordinates, using $(\mathrm{SEM}_T/T)^{-2}$ as weights. Exponent uncertainties are one-standard-error covariance estimates, multiplied by $\sqrt{\chi_\nu^2}$ when the reduced chi-squared exceeded unity.

The inset is a separate qualitative matched-driving comparison and was not used for the main-panel calibration. It uses the original five-realization finite-difference analysis in reduced units, evaluated over the final $20\%$ of active-cluster and final $50\%$ of passive-cluster trajectories, and serves only to show that passive clusters rotate more slowly at matched driving magnitude and particle-size ratio.

\subsection*{Edge-current analysis}
\label{sec:edge-current-analysis}

To distinguish boundary-localized currents from rigid-body cluster rotation, we determined the translational velocity $\vec V$, angular velocity $\Omega$, and center $\vec R$ of each cluster. The translation-corrected tangential velocity was calculated as
\begin{equation*}
    v_{\theta,i}
    =
    \frac{
        \hat{\vec z}\cdot
        \left[
            (\vec r_i-\vec R)
            \times
            (\vec v_i-\vec V)
        \right]
    }{
        |\vec r_i-\vec R|
    }.
\end{equation*}
For ideal rigid-body rotation, $v_\theta(r)=\Omega r$. Edge currents were therefore identified from systematic deviations of the measured profile $v_\theta(r)$ from this rigid-body reference. The corresponding tangential velocity in the co-rotating frame is $u_\theta=v_\theta-\Omega r$.

\paragraph{Experiments}

Experimental velocity fields were obtained from image sequences of isolated particle clusters using ensemble-averaged particle-image velocimetry in PIVlab \cite{thielickeParticle2021}. The drift of the particle cluster was also corrected using the center of mass of the cluster. Only valid PIV vectors located within the selected isolated cluster were retained. The cluster center and angular velocity were determined by fitting the velocity field in the central region to
\begin{equation*}
    \vec v(\vec r)
    =
    \vec V
    +
    \Omega\,
    \hat{\vec z}\times(\vec r-\vec R).
\end{equation*}
The azimuthal component was then evaluated using the expression for $v_{\theta,i}$ mentioned above and averaged in radial bins about $\vec R$. The rigid-body reference shown in Fig.~\ref{fig:edge-currents} was obtained from a linear fit through the origin to the innermost five points of the radial profile.

\paragraph{Simulations}

For Fig.~\ref{fig:edge-currents}a,b, a cluster of about \SI{1e4}{} passive particles was initialized on a hexagonal lattice at a local covered-area fraction of approximately $0.73$. Active particles were distributed randomly in the surrounding bath, giving $N=57,600$, $\phi_{\mathrm P}=0.2$, and a total covered-area fraction $\phi=0.5$. Following energy minimization and $10^{5}$ WCA-only equilibration steps, production trajectories were generated for $10^{8}$ time steps and recorded every $5{,}000$ steps. The magnetic attraction strength and the depletion strength were chosen to $\beta=0.5$ and $\alpha_{\mathrm{pp}}=2.8$, respectively.

For Fig.~\ref{fig:edge-currents}d, we performed simulations of a passive cluster with 80 particles at $\phi_\mathrm{P}=0.2$, $\beta=0.2$, and $\alpha_{\mathrm{pp}}=5.0$ for $10^{8}$ time steps and recorded every $5{,}000$. The simulation data was scaled to physical units by fitting the conversion factor such that the simulated cluster's outer radius matches the experimentally observed cluster radius \SI{11.0}{\um}, since the fixed particle-size ratio of $1.2$ used in the simulation differs from the experimental one \(\SI{2.1}{\um} / \SI{1.2}{\um} \approx 1.75\). This scaling is intended to demonstrate qualitative agreement in the edge-current profile, not a quantitative prediction from independently measured particle sizes.

Passive particles belonging to the largest cluster were identified in each frame using a distance-based connectivity criterion with a cutoff of $1.5\sigma_{\mathrm p}$, accounting for periodic boundary conditions. The cluster center was calculated from the circular mean of the particle coordinates in each periodic direction. Particle velocities were obtained from displacements between consecutive stored configurations.

Translation and rigid-body rotation were simultaneously fitted using particles within a persistent cluster core. The core comprised particles lying within the innermost $60\%$ of the radial distribution in at least $80\%$ of the analyzed frames. A robust least-squares fit of
\begin{equation*}
    \vec v_i
    =
    \vec V
    +
    \Omega\,
    \hat{\vec z}\times(\vec r_i-\vec R)
    +
    \vec u_i
\end{equation*}
yielded $\vec V$, $\Omega$, and the residual velocity $\vec u_i$. The translation-corrected tangential velocity was averaged over trailing windows of $50\tau_{\mathrm B}$ and subsequently binned by the radial distance $r=|\vec r_i-\vec R|$. We used 28 equally spaced radial bins extending to the $99.5$th percentile of the sampled cluster radii. The dashed line in Fig.~\ref{fig:edge-currents} shows the corresponding rigid-body profile $|\Omega|r$ fitted to the average angular velocity of the cluster core.


\bmhead{Acknowledgements}
The authors thank the Nanobiophysics and Physics of Complex Fluids groups (University of Twente, the Netherlands) for providing access to their laboratory facilities. The authors also thank C.J. Padberg and I. Punt for SEM measurements. The authors gratefully acknowledge the computing time provided to them on the high-performance computer Lichtenberg II at TU Darmstadt, funded by the German Federal Ministry of Research, Technology and Space (BMFTR) and the State of Hesse.

\bmhead{Funding}
H.R.V. acknowledges funding from the Netherlands Organisation for Scientific Research (NWO-M1-OCENW.M.21.309) and the European Research Council (ERC Consolidator Grant no. 101171050-SynthAct3D).
D.S. and B.L. acknowledge funding from the Deutsche Forschungsgemeinschaft through the FOR5584 (Project No. 509491635).

\bmhead{Author contributions}
H.R.V. and B.L. conceived and designed the experiments and the simulation model, respectively. D.S. performed the simulations. A.N.S. and S.v.d.H. performed the experiments. D.S., A.N.S. and S.v.d.H. analyzed the data. B.L. and H.R.V. supervised the project. All authors were involved in writing and editing the manuscript.

\bmhead{Competing Interests}
All the authors declare no competing interests.

\bmhead{Data, materials and code availability}
Data and code will be deposited in a repository. Materials can be provided upon reasonable request.

\bibliography{state-inversion-v2}

\end{document}